\documentclass[prl,aps,letterpaper,floatfix,superscriptaddress,nofootinbib,preprintnumbers,twocolumn,10pt]{revtex4-1}

\usepackage{dcolumn}
\usepackage{bm}
\usepackage[dvips]{graphicx}
\usepackage{epsfig}
\usepackage{slashed}
\usepackage{hhline}
\usepackage{amsmath, amsfonts, amssymb}
\usepackage[normalem]{ulem}
\newcommand{\be}{\begin{equation}}
\newcommand{\ee}{\end{equation}}
\newcommand{\bee}{\begin{equation*}}
\newcommand{\eee}{\end{equation*}}
\newcommand{\bea}{\begin{eqnarray}}
\newcommand{\eea}{\end{eqnarray}}
\newcommand{\bean}{\begin{eqnarray*}}
\newcommand{\eean}{\end{eqnarray*}}

\usepackage{color}

\begin{document}
\preprint{KCL-PH-TH/2017-28}

\title{Probing the Pseudoscalar portal to Dark Matter via $\bar{b} b Z( \rightarrow \ell\ell) + \slashed{E}_{T}$:\\
From the LHC to the Galactic Centre Excess}


\author{Patrick Tunney} \email{patrick.tunney@kcl.ac.uk}
\affiliation{Department of Physics, King's College London, Strand, WC2R 2LS London, UK}

\author{Jose Miguel No} \email{jose\_miguel.no@kcl.ac.uk}
\affiliation{Department of Physics, King's College London, Strand, WC2R 2LS London, UK}

\author{Malcolm Fairbairn} \email{malcolm.fairbairn@kcl.ac.uk}
\affiliation{Department of Physics, King's College London, Strand, WC2R 2LS London, UK}


\date{\today}

\begin{abstract}
We propose a new search for Dark Matter at the LHC, characteristic of scenarios beyond the Standard Model with a 
pseudoscalar portal between the visible and dark sectors. 
This search, leading to a $\bar{b} b Z( \rightarrow \ell\ell) + \slashed{E}_{T}$ final state, 
reaches large regions of parameter space not probed by Dark Matter searches via multi-jet $+ \slashed{E}_{T}$, 
searches for new scalars and flavour bounds. We show that this search could be used to test   
the Dark Matter origin of the gamma ray Galactic Centre excess with LHC Run 2 data. 
\end{abstract}

\maketitle



\subsection*{I. Introduction}

\vspace{-4mm}

The nature of dark matter (DM) is an unsolved mystery at the interface of particle physics and cosmology. 
One widely studied DM candidate is the Weakly-Interacting-Massive-Particle (WIMP), whose relic abundance 
is obtained via thermal freeze-out in the early Universe with a mass in the range 
$\mathrm{GeV}-\mathrm{TeV}$ (see~\cite{Bertone:2004pz} for a review).


There is an ongoing multi-pronged experimental effort to search for WIMP DM via its interactions with Standard Model particles: indirectly 
by measuring the energetic particles produced by DM annihilations in space and directly by measuring the scattering of ambient DM from nuclei. 
The observed gamma ray excess in the Fermi-LAT space telescope observations of the Milky Way Galactic Centre~\cite{TheFermi-LAT:2015kwa} 
may be interpreted as the existence of weak-scale DM annihilating 
into $\bar{b} b$ pairs~\cite{Goodenough:2009gk,Hooper:2010mq,Hooper:2011ti,Abazajian:2012pn} (see~\cite{Karwin:2016tsw} for a recent exhaustive
analysis of the excess and its DM interpretation). 
While arguably there is some tension between the DM interpretation of the gamma ray excess at the Galactic Centre and the 
non-observation of emission due to DM annihilation in dwarf 
spheroidal galaxies \cite{Ahnen:2016qkx}\footnote{However we are also aware that the errors on the astrophysical J-factors used in \cite{Ahnen:2016qkx} are 
somewhat small and allowing more freedom in the fit and adding a systematic error representing the possibility of triaxiality in the halos could reduce this 
disagreement somewhat \cite{Bonnivard:2015xpq}.}, the self-annihilation cross section needed to explain the excess can be consistent with that required to generate the 
observed relic abundance through thermal freeze-out in the early 
Universe $\langle\sigma \mathrm{v}\rangle  \simeq 3\times 10^{-26}{\rm  cm}^3/{\rm s}$.~At the same time, 
current limits on the spin-independent DM interaction 
cross section with nuclei by the Large-Underground-Xenon (LUX)~\cite{Akerib:2016vxi} and PandaX~\cite{Tan:2016zwf} experiments strongly constrain 
DM masses in the range $10-100$ GeV. A compelling DM interpretation of the gamma ray Galactic Centre excess (GCE) in combination 
with the non-observation of a signal in DM direct detection experiments is via the existence of a pseudoscalar mediator between 
the visible and DM sectors~\cite{Boehm:2014hva,Izaguirre:2014vva,Ipek:2014gua}, which yields  
spin-dependent DM-nucleon interactions, for which experimental limits are much less stringent.~Pseudoscalar 
mediated DM-nucleon interactions generally lie well below the reach of present DM direct detection experiments.


Direct and indirect probes of DM are complemented by searches at colliders, where pairs of DM particles could be produced. These escape the detector and 
manifest themselves as events possessing an imbalance in momentum conservation, via the presence of missing transverse 
momentum $\slashed{E}_{T}$ recoiling against 
a visible final state $X$. Searches for events with large $\slashed{E}_{T}$ are currently a major focus at the Large Hadron Collider (LHC) largely
due to their connection to DM~\cite{Morrissey:2009tf}. 
%
In this work we present a new search avenue for DM at the LHC, characteristic of renormalizable, gauge invariant scenarios beyond the Standard Model with a 
pseudoscalar portal between the visible and dark sectors. The search is characterized by 
a $b\bar{b} \,Z \,\,(Z \to \ell\ell) + \slashed{E}_{T}$ final state. We show that this new 
DM search channel of a leptonically decaying $Z$ boson, two bottom quarks and missing transverse momentum
will yield a powerful probe of the region of parameter space consistent with a DM interpretation of the GCE through
LHC Run 2 data. 

\vspace{-4mm}

\subsection*{II. The Pseudoscalar Portal Into Dark Matter}

\vspace{-4mm}

We focus our analysis on scenarios with a pseudoscalar mediator between DM and the SM fermions. These can yield a compelling explanation of the 
GCE through DM annihilation into $b$-quarks (see e.g.~\cite{Izaguirre:2014vva,Ipek:2014gua}).
For concreteness we consider DM to be a Dirac fermion $\chi$ with mass $m_{\chi}$, singlet under the SM gauge interactions and  
coupling to a real singlet pseudoscalar mediator $a_0$ via 
\begin{equation}
\label{Ldark}
 V_{\mathrm{dark}} = \frac{m^2_{a_0}}{2}\,a_0^2 + m_{\chi}\, \bar{\chi}\chi + y_{\chi}\,a_0 \,\bar{\chi} i\gamma^{5} \chi\, .
\end{equation}
However, for the pseudoscalar to be able to mediate interactions between DM and the SM 
fermions, $SU(2)_{\mathrm{L}} \times U(1)_{\mathrm{Y}}$ gauge invariance requires the existence of new states beyond the SM
in addition to the DM particle and the pseudoscalar mediator~\cite{Nomura:2008ru,Goncalves:2016iyg}. 
A renormalizable and gauge invariant realization of the pseudoscalar portal between DM and the SM leads to the extension of  
the SM Higgs sector with a second Higgs doublet, as first 
noted in~\cite{Nomura:2008ru}.~A theory 
with the required ingredients then naturally resembles a two Higgs doublet model (2HDM)~\cite{Nomura:2008ru,Goncalves:2016iyg,No:2015xqa,Bauer:2017ota}.
We note that this also yields a compelling explanation for the 
preferential coupling of the pseudoscalar mediator to third generation SM fermions ($b$-quarks and $\tau$-leptons), in relation to the GCE. 
 
In the following we provide a brief review of the 2HDM aspects of relevance to us 
(for a general review of 2HDM theory and phenomenology, see e.g.~\cite{Branco:2011iw}):
The two Higgs doublets are $H_j = \left(\phi_j^{+} , (v_j + h_j + i \, \eta_j)/\sqrt{2} \right)^T$, with ($j = 1,2$). $v_j$ are the \textit{vev} of the 
doublets ($\sqrt{v^2_1 + v^2_2} = v$ and $v_2/v_1 \equiv \mathrm{tan} \beta$).
We consider a 2HDM scalar potential with 
%
Charge-Parity (CP) conservation and a softly broken $\mathbb{Z}_2$ symmetry. 
The presence of this $\mathbb{Z}_2$ symmetry in the couplings of the doublets $H_j$ to fermions allows to forbid dangerous 
tree-level flavour changing neutral currents, by forcing each fermion type to couple to one doublet 
only~\cite{Glashow:1976nt}.~In 
{\sl Type I} 2HDM all fermions couple to $H_{2}$, while for {\sl Type II} 2HDM up-type quarks couple to $H_{2}$ and down-type quarks and leptons couple to $H_{1}$.
The scalar spectrum of the 2HDM contains a charged scalar $H^{\pm} = \mathrm{cos}\beta \,\phi_2^{\pm} - \mathrm{sin}\beta \, \phi_1^{\pm}$, a 
neutral CP-odd scalar $A_0 = \mathrm{cos}\beta \,\eta_2 - \mathrm{sin}\beta \, \eta_1$
and two neutral CP-even scalars $h = \mathrm{cos}\alpha \,h_2 - \mathrm{sin}\alpha \, h_1$, $H_0 = - \mathrm{sin}\alpha \,h_2 - \mathrm{cos}\alpha \, h_1$.
We identify $h$ with the 125 GeV Higgs state, which has SM-like properties when the mixing angle $\alpha$ in the neutral CP-even sector satisfies 
$\beta - \alpha = \pi/2$. 

As we show now, the 2HDM allows for pseudoscalar mediated interactions between the visible sector and the DM candidate $\chi$ in~\eqref{Ldark}. 
The portal between the visible and dark sectors occurs via
\begin{equation}
\label{Vportal}
V_{\mathrm{portal}} = i\,\kappa\,a_0 \,H_1^{\dagger}H_2 + \mathrm{h.c.}
\end{equation}
which causes the would-be 2HDM state $A_0$ to mix with $a_0$, 
yielding two pseudoscalar mass eigenstates $a,A$: $a = c_{\theta} \,a_0 - s_{\theta} \, A_0$, $A = c_{\theta} \,A_0 + s_{\theta} \, a_0$, 
with $c_{\theta} \equiv \mathrm{cos}\theta$ and $s_{\theta} \equiv \mathrm{sin}\theta$. 
This mixing allows both $a$ and $A$ to couple simultaneously to DM and the SM fermions, providing the portal between visible and DM sectors.
The coupling of $a$ $(A)$ to DM is given by $s_\theta\, y_{\chi}$ ($c_\theta\, y_{\chi}$). Regarding the 
pseudoscalar couplings to SM fermions, these are given by $g_{\mathrm{SM}}\times y_{f}/\sqrt{2}$ 
(with $y_{f}$ the Yukawa coupling of the fermion). We consider here a Type II 2HDM, for which the 
$g_{\mathrm{SM}}$ coupling of $a$ $(A)$ is given by $s_{\theta}\,\mathrm{tan}^{-1}\beta$ 
($c_\theta\, \mathrm{tan}^{-1}\beta$) for up-type quarks and
$s_\theta\, \mathrm{tan}\beta$ ($c_\theta\, \mathrm{tan}\beta$) for down-type quarks and charged leptons.
To simplify the following discussion, we also restrict ourselves to $\beta - \alpha = \pi/2$ (the so-called alignment limit) 
where $h$ behaves exactly as the SM Higgs~\cite{Gunion:2002zf}.
We note that for a Type II 2HDM, deviations from the alignment limit are strongly constrained by LHC Higgs 
measurements~\cite{Aad:2015pla}.

\vspace{1mm}

For the rest of this work, we consider
the benchmark value $m_{\chi} = 45$ GeV: For a pseudoscalar mediator,~\cite{Karwin:2016tsw} finds a preferred range $m_{\chi} \in [50,\,170]$ GeV
if DM annihilates into $b$-quark pairs, and $m_{\chi} \in [10,\,20]$ GeV if it annihilates into leptons, concerning the GCE. 
In the present case, DM annihilates dominantly into $b$-quarks, with a small ($\sim 10\,\%$) annihilation component into 
$\tau$-leptons.
We also consider the mediator $A$ (doublet-like) to be much heavier than $a$ (singlet-like). 
For $m_{\chi} < m_a \ll m_A$, DM annihilates to SM particles through $s$-channel $a$ exchange. 
The velocity averaged annihilation cross section for $\chi\bar\chi\to{\rm SM}$ in the nonrelativistic limit is
\begin{align}
\langle \sigma \mathrm{v}\rangle&=\frac{y_\chi^2}{2\pi}\frac{m_\chi^2}{m_a^4}s_{\theta}^2 c_{\theta}^2\,
\mathrm{tan}^2\beta\left[\left(1-\frac{4m_\chi^2}{m_a^2}\right)^2+\frac{\Gamma_a^2}{m_a^2}\right]^{-1}
\nonumber
\\
&\quad\quad\quad\quad\times\sum_{f} N_C \frac{m_f^2}{v^2}\sqrt{1-\frac{m_f^2}{m_a^2}}.
\label{eq:annih}
\end{align}
with $\Gamma_a$ the decay width of $a$. The sum is over quarks ($N_C=3$) and charged 
leptons ($N_C=1$). Reproducing the observed DM relic density requires $\langle \sigma \mathrm{v}\rangle \simeq 3\times 10^{-26}{\rm  cm}^3/{\rm s}$,
which favours large values of $\mathrm{tan}\beta$ (particularly for not too large values of $y_{\chi}$). 

\vspace{2mm}

Flavour constraints from $\bar{B}\to X_s \gamma$ decays yield a lower bound on $m_{H^{\pm}}$ in Type II 2HDM, 
given by $m_{H^{\pm}} > 480$ GeV at 95\% C.L.~\cite{Misiak:2015xwa}. In addition, electroweak precision observables 
strongly constrain the splitting between the charged scalar $H^{\pm}$ and either of the neutral states 
$H_0$, $A$~\cite{Grimus:2007if}.~Combined, these yield $m_{A},\, m_{H_0}, \,m_{H^\pm} \gtrsim 500$ GeV. 
On the other hand $m_{A},\, m_{H_0}, \,m_{H^\pm}$ may not be taken arbitrarily high if  
$s_{\theta}$ and/or $m_a$ are kept fixed due to unitarity constraints.
For $m_a \sim 100$ GeV and $\theta = \pi/4$ the unitarity bounds on $m_{A},\, m_{H_0}$ are respectively 
$m_{A} \leq 1.4$ TeV, $m_{H_0} \leq 1$ TeV~\cite{Goncalves:2016iyg}.
In the following we take as benchmarks $m_{H^{\pm}} = m_{H_0} = 600$ GeV, $800$ GeV (and assume a somewhat larger $m_A$).

\vspace{-5mm}

\subsection*{III. Experimental Constraints on the Dark Portal}

\vspace{-3mm}

The above pseudoscalar dark portal scenario is constrained in a variety of ways.
Besides the aforementioned flavour bound $m_{H^{\pm}} > 480$ GeV from $\bar{B}\to X_s \gamma$ decays, the existence of a 
light pseudoscalar $a$ coupling to SM fermions can be probed by its contributions to the 
decay $B_s\to\mu^+\mu^-$~\cite{Skiba:1992mg,Logan:2000iv}, which for $m_a \gg m_{B_s} \simeq 5.36$ GeV may be 
expressed as\footnote{We note the important $H^{\pm}$ 
contribution in the limit $s_{\theta} \to 0$ (see~\cite{Logan:2000iv}) which was missed in~\cite{Ipek:2014gua}.}
\begin{align}
&\mathrm{Br}\left(B_s\to\mu^+\mu^-\right)\simeq\mathrm{Br}\left(B_s\to\mu^+\mu^-\right)_{\rm SM}
\\
&\times\left(\left|1+ x_B \mathrm{tan}^2\beta \frac{f(x_t,x_a,r)}{4 \,Y(x_t)}\right|^2+ 
\left|x_B \mathrm{tan}^2\beta\frac{g(r)}{4 \,Y(x_t)}\right|^2\right),
\nonumber
\end{align}
with $x_B=m_b m_{B_s}/m_W^2$, $x_t=m_t^2/m_W^2$, $x_a=m_a^2/m_A^2$, $r=m_{H^\pm}^2/m_t^2$, $r_t=x_t\,r$, $g(r) = \log(r)/(r-1)$,
\begin{eqnarray}
 f\left(x_t,x_a,r\right)= g(r) &+& \frac{s^2_{\theta}}{(r-1)}\left[ 2c^2_{\theta}(x_a + x_a^{-1} -1)-1\right]\nonumber \\
 &\times & \left( \frac{r_t \log r_t}{\left(r_t-1\right)} - \frac{x_t \log x_t}{\left(x_t-1\right)} \right)\,,
 \end{eqnarray}
and $Y(x)$ the Inami-Lim function,
\begin{align}
Y(x)&=\frac{x}{8 (x-1)^2}\left[4 - 5 x +x^2 + 3 x \log x\right].
\end{align}
The average of the LHCb and CMS measurements of this mode from LHC 7 and 8 TeV data is
$\mathrm{Br}\left(B_s\to\mu^+\mu^-\right)=\left(2.9\pm0.7\right)\times10^{-9}$~\cite{Aaij:2013aka,Chatrchyan:2013bka,CMSandLHCbCollaborations:2013pla} 
which may be compared against the SM prediction
$\left(3.65\pm0.23\right)\times10^{-9}$~\cite{Bobeth:2013uxa,Buras:2013uqa}.

For $m_a < m_h/2$ the presence of the decay $h \to a a$ yields stringent constraints on the model~\cite{Ipek:2014gua}, and consequently we only consider 
here the case $m_a > m_h/2$ for which non-standard Higgs decays are suppressed (note that for $m_{\chi} = 45$ GeV the 3-body decay $h \to a \bar{\chi}\chi$ 
is also kinematically forbidden above $m_a = 35$ GeV).

LHC searches for the states $H_0$, $A$ and $a$ decaying to $\tau^+\tau^-$ also place important constraints at large $\mathrm{tan}\beta$
($a \to \bar{b}b$ has also been considered, see e.g.~\cite{Kozaczuk:2015bea}).
Focusing on $\phi = H_0, \,a$, the latest CMS search for $\bar{b}b \phi \,(\phi\to\tau^+\tau^-)$ 
with an integrated luminosity of 12.9 fb$^{-1}$~\cite{CMS:2016rjp} yields limits on the parameter space for 
$m_a$, $m_{H_0}$, $s_{\theta}$, $\mathrm{tan}\beta$. 

Finally, the pseudoscalar portal to DM can be probed at the LHC in the $\bar{t}t + \slashed{E}_{T}$ and $\bar{b}b + \slashed{E}_{T}$ channels 
(see~\cite{Banerjee:2017wxi} for a recent discussion), and in multi-jet $ + \slashed{E}_{T}$~\cite{Buchmueller:2015eea}. 
Using the results from~\cite{CMS:2016uxr} we find that $\bar{b}b + \slashed{E}_{T}$ searches at $\mathrm{tan}\beta \gg 1$ yield
significantly weaker constraints that the ones discussed above (e.g. $B_s\to\mu^+\mu^-$). At the same time, 
$\bar{t}t + \slashed{E}_{T}$ searches are currently only sensitive to $\mathrm{tan}\beta < 1$.
For multi-jet $ + \slashed{E}_{T}$ searches, using the analysis from~\cite{Buchmueller:2015eea} we find that 
these yield an important constraint at low $\mathrm{tan}\beta$, but still being subdominant to those from the 
searches discussed in the next Section. 

\vspace{-4mm}

\subsection*{IV. A New LHC Probe of Dark Matter}

\vspace{-4mm}

Remarkably, when $m_{H_0} \gg m_a$ the decay $H_0 \to Z a$ yields a new avenue to probe DM at the 
LHC.~For $\mathrm{tan}\beta \gg 1$ as favoured by the GCE, a novel DM search channel presents itself: 
$p p \to b\bar{b}\, H_{0}, \, H_0 \to Z \,a$ $(Z \to \ell^+ \ell^-, \, a \to \bar{\chi} \chi )$. This topology for the 
final state $\bar{b} b \, \ell^+ \ell^- + \slashed{E}_{T}$ has not yet 
been explored at the LHC, and we show here that this signature allows to probe a wide range of 
parameter space for pseudoscalar portal scenarios, in particular within the region consistent with a DM interpretation of the GCE.

In order to study the prospects for this signature at the LHC with $\sqrt{s} = 13$ TeV c.o.m.~energy, 
we require events with two oppositely charged electrons/muons in the invariant mass window $m_{\ell\ell} \in \left[76,106\right]$ GeV, 
with $p^{\ell}_T > 20$ GeV and rapidity $|\eta^{\ell}| < 2.5$. Our event selection further requires
$|p_T^{\ell\ell} - \slashed{E}_{T}|/p_T^{\ell\ell} < 0.5$ and a separation $\Delta R_{\ell \ell} > 0.4$ between the 
same-flavor lepton pair. We also demand at least one $b$-tagged jet with\footnote{We note that 
a very low value of the chosen $p^{b}_T$ cut (for a very high value of $m_{H_0}$) 
could result in a breakdown of the perturbative expansion~\cite{Degrande:2016aje} for the $\bar{b}b$-associated production 
of $H_0$ (we thank Richard Ruiz for pointing out this issue to 
us).~Using {\sc SusHi}~\cite{Harlander:2012pb} We have estimated our $b\bar{b}H_0$ next-to-leading-order (NLO) $k$-factor to be $\sim 1.4$, 
close to the perturbative expansion validity limit, but arguably safe~\cite{Degrande:2016aje}.
} $p^{b}_T > 30$ GeV.

\begin{figure}[h!]
\begin{center}
\includegraphics[width=0.486\textwidth]{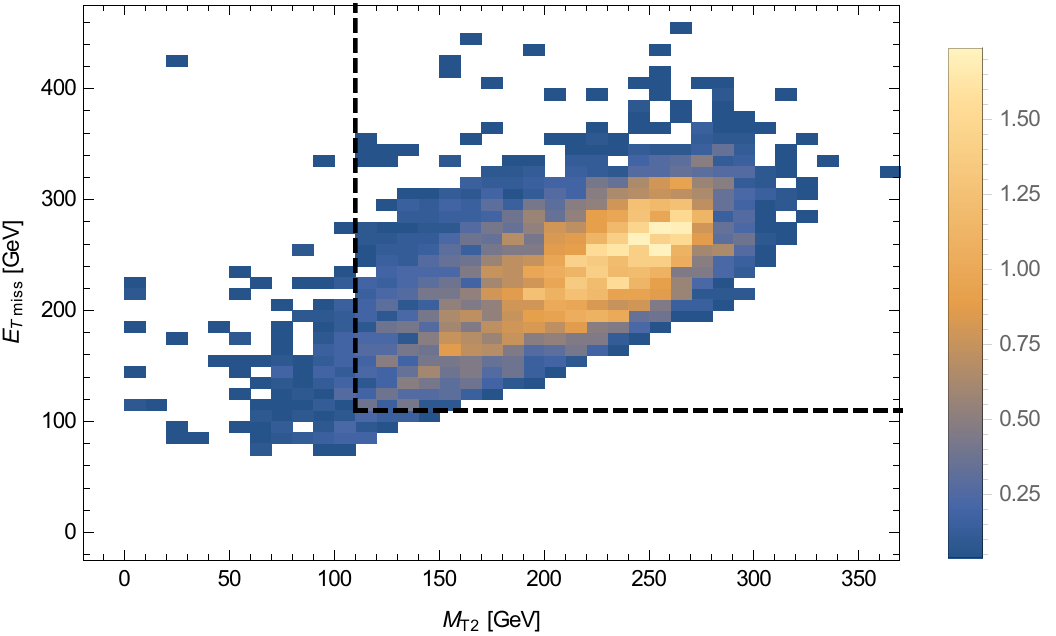}

\includegraphics[width=0.485\textwidth]{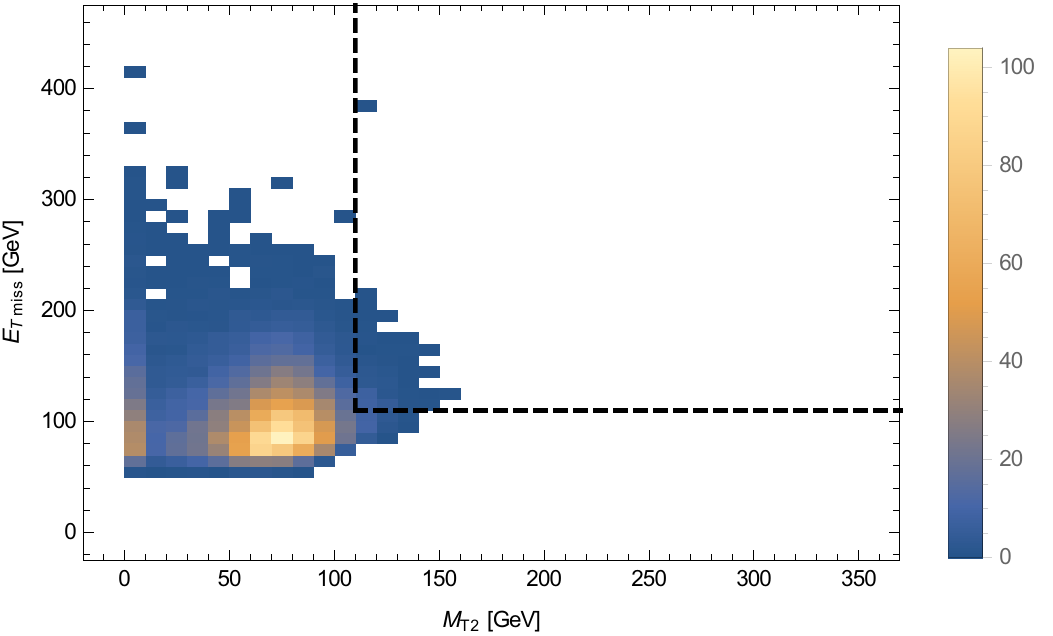}
\caption{\small $m_{H_0} = 600$ GeV, $m_a = 150$ GeV signal (top) and $t\bar{t}$ background (bottom) events after 
event selection with $300$ fb$^{-1}$, in the ($m_{T2}$, $\slashed{E}_T$) plane. The dashed line corresponds to 
the signal region $\slashed{E}_{T},\, m_{T2} > 110$ GeV.
}
\label{fig:2}
\end{center}

\vspace{-2mm}

\end{figure}

The main SM backgrounds are $t \bar{t}$ and di-boson ($WZ$ and $ZZ$) + jets production\footnote{The $W W$ diboson background is strongly suppressed by the 
$m_{\ell\ell}$ selection in combination with a large amount of $\slashed{E}_{T}$. 
Other potential backgrounds become negligible when demanding 
a significant amount of $\slashed{E}_{T}$ in the event.}.
The requirement of one or more $b$-tagged jets acts as an effective suppressor of the latter, while the invariant mass window $m_{\ell\ell}$
helps diminish the $t \bar{t}$ background. In order to further reduce SM backgrounds we take advantage of the boosted configuration of the signal 
for $m_{H_0} \gg m_a + m_Z$, and demand the leading lepton in $p^{\ell}_T$ to satisfy $p^{\ell_1}_T > 80$ GeV as well as 
$p^{\ell_1}_T + p^{\ell_2}_T > 150$ GeV. Finally we use $\slashed{E}_{T}$ and the variable 
$m_{T2}$  \cite{Lester:1999tx} to define our signal region. 
We calculate $m_{T2}$ using  \cite{Lester:2014yga} as
\begin{align}
 \label{eq:mT2}
\hspace{-2mm} m^2_{T2} \equiv \min_{\scriptscriptstyle{ \vec{k_T} + \vec{q_T} = \slashed{\vec{p_T}}}} \left\lbrace \max 
\left[ m_T^2(\vec{p_T^{\ell^+}},\vec{k_T}), m_T^2 (\vec{p_T^{\ell^-}},\vec{q_T}) \right] 
\right\rbrace
\end{align}
where minimisation is over all possible vectors $\vec{k_T}$ and $\vec{q_T}$ that satisfy $\vec{k_T} + \vec{q_T} = \slashed{\vec{p_T}}$
(with $\left|\slashed{\vec{p_T}}\right| = \slashed{E}_{T}$). 
Our signal region is defined as\footnote{The $m_{T2}$ cut is chosen conservatively to ensure the background prediction is 
not dominated by the Monte Carlo statistical uncertainty. An analysis performed by the experimental collaborations would achieve 
better sensitivity through a stronger cut on $m_{T2}$.}  $\slashed{E}_{T} > 110$ GeV, $m_{T2} > 110$ GeV.

We generate our signal and background event samples at LO in {\sc MadGraph5$\_$MC@NLO}~\cite{Alwall:2014hca} and perform 
showering in Pythia 8~\cite{Sjostrand:2014zea}. 
For the $Z Z$ and $W Z$ backgrounds we include up to two additional jets in the final state, matched to parton shower.
%
%
%
%
%
We replace a full detector simulation with a Gaussian smearing of the $p_T$ of final state paricles: We define jets, well isolated charged leptons and photons, 
and $\slashed{E}_{T}$ as the relevant final state objects. Jets are constructed with the FastJet package~\cite{Cacciari:2011ma} using the anti-$k_T$ 
algorithm~\cite{Cacciari:2008gp} with $R =0.4$, and are required to have $p_T > 25$ GeV and $| \eta | < 2.5$. We smear the $p_T$ of the visible 
particles and calculate both the truth $\slashed{E}_{T}$ and the reconstructed value calculated from the smeared visible objects. We then 
smear the difference between the truth and reconstructed $\slashed{E}_{T}$. The functions for the smearing of the visible 
objects and $\slashed{E}_T$, as well as the b-tagging efficiency and mistag rates, are chosen to match the ATLAS performance 
reported in~\cite{Aad:2009wy} for the leptons and $\slashed{E}_T$,~\cite{Aad:2012ag} for the jets and~\cite{b-tag} for the b-tagging.
We derive the projected sensitivity of our search using the CLs method~\cite{Read:2002hq}, and  
assuming a conservative $20\%$ background systematic uncertainty added in quadrature to a $1/\sqrt{N}$ Monte Carlo uncertainty
($N$ the number of generated background Monte Carlo events in the signal region). 

For a benchmark signal $m_{H_0} = 600$ GeV, 
$m_a = 150$ GeV, $\mathrm{tan}\beta = 15$, $s_{\theta} = 0.3$ the background and signal samples 
surviving event selection are shown 
in Figure~\ref{fig:2} in the ($\slashed{E}_T$, $m_{T2}$) plane, highlighting the choice of signal region  
$\slashed{E}_{T},\,m_{T2} > 110$ GeV as tailored for a clean signal extraction.
In Figure~\ref{PLOT1} we show the 95\% C.L. sensitivity of our proposed search (hatched region)
with $300$ fb$^{-1}$ of integrated luminosity
in the ($s_{\theta},\,\mathrm{tan}\beta$) plane for ($m_{H_0},\,m_a$) = ($600,\,150$) GeV (left) and 
($800,\,150$) GeV (right), demanding 
$\langle \sigma \mathrm{v}\rangle \simeq 3\times 10^{-26}{\rm  cm}^3/{\rm s}$ to fix 
$y_{\chi}$ in terms of $\mathrm{tan}\beta$ 
and $s_{\theta}$ in each case. We demand perturbativity $y_{\chi} < 4 \pi$, and show the lines $y_{\chi} = 1$ (dotted grey) and $y_{\chi} = 0.1$ (dot-dashed grey) 
for guidance.

\begin{widetext}

\vspace{-1mm}
\onecolumngrid

\begin{figure}[ht!]
\begin{center}

$\vcenter{\hbox{\includegraphics[width=0.80\textwidth]{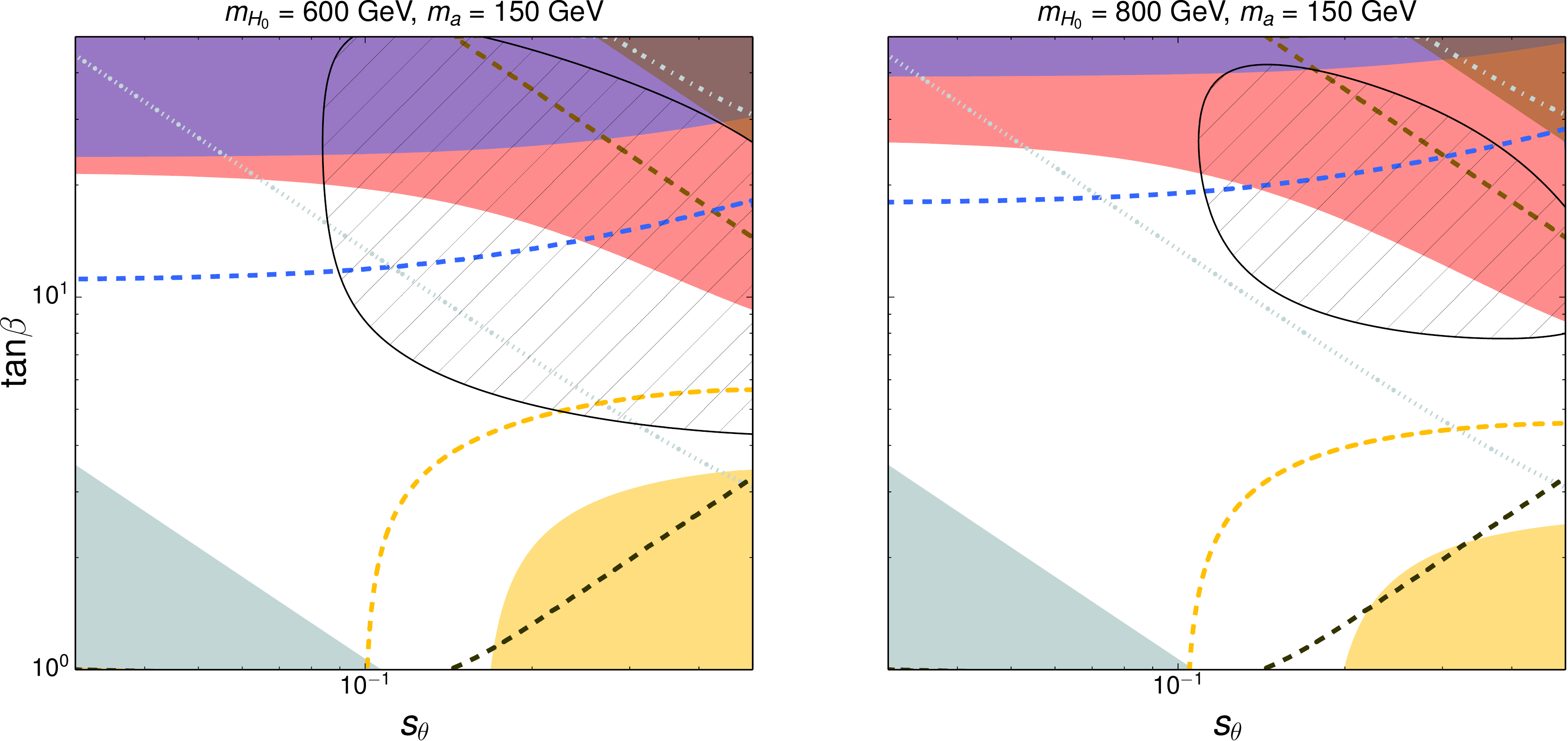}}}$
\hspace{1mm}
$\vcenter{\hbox{\includegraphics[width=0.18\textwidth]{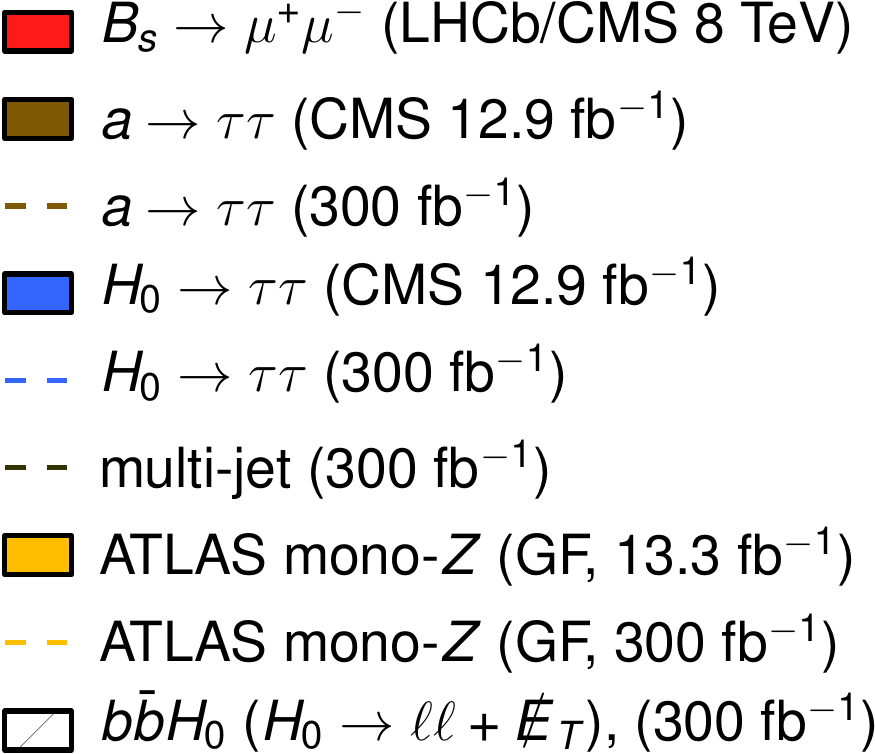}}}$

\caption{\small Current (solid) and projected LHC with $\mathcal{L}=300$ fb$^{-1}$ (dashed lines)  
95\% C.L. exclusion regions in the ($s_{\theta},\,\mathrm{tan}\beta$) plane for ($m_{H_0},\,m_a$) = ($600,\,150$) GeV (left) and 
($800,\,150$) GeV (right) with a DM mass $m_{\chi} = 45$ GeV, from $H_0 \to \tau^+ \tau^-$ (blue), $a \to \tau^+ \tau^-$ (brown), 
multi-jet $+ \slashed{E}_T$ (black)
and ATLAS mono-$Z$ GF (yellow). 
Exclusion from CMS/LHCb 8 TeV
$B_s\to\mu^+\mu^-$ measurements is shown in red. 
The dashed region corresponds to the 
95\% C.L. sensitivity for our proposed search, $pp\rightarrow b\bar{b}\,\ell\ell+\slashed{E_T}$, with $\mathcal{L}=300$ fb$^{-1}$.
The coupling $y_{\chi}$ is fixed at each point to match the DM relic density.
The perturbativity excluded region $y_{\chi} > 4 \pi$ is depicted in grey.  
Lines $y_{\chi} = 1$ (dotted grey), $y_{\chi} = 0.1$ (dot-dashed grey) are shown for guidance.}
\label{PLOT1}
\end{center}
\end{figure} 

\vspace{-6mm}
\end{widetext}

The decay $H_0 \to Z a$ ($a\to \bar{\chi} \chi$) may be probed also by ATLAS/CMS mono-$Z$ searches in the $\ell^+\ell^- + \slashed{E}_{T}$ 
channel~\cite{ATLAS-CONF-2016-056,CMS-PAS-EXO-16-038}, both for gluon-fusion (GF) production of $H_0$ and for $b\bar{b}$-associated production 
(if both $b$-jets are missed, since~\cite{ATLAS-CONF-2016-056,CMS-PAS-EXO-16-038} impose jet/$b$-jet vetoes).  
We follow the LHC 13 TeV analysis selection of ATLAS~\cite{ATLAS-CONF-2016-056} with $13.3\,\mathrm{fb}^{-1}$ to derive present 95\% C.L. constraints   
on our signal in the ($s_{\theta}$, $\mathrm{tan}\beta$) plane, 
shown in Figure~\ref{PLOT1} for GF (yellow region) for 
$m_{H_0} = 600$ GeV, $m_a = 150$ GeV (Left) and $m_{H_0} = 800$ GeV, $m_a = 150$ GeV (Right).  
We also show the LHC 
projections to $300$ fb$^{-1}$ (dashed lines) using a naive~$\sqrt{\mathcal{L}}$ increase in the signal cross section sensitivity 
(we note that even in this case, the ATLAS mono-$Z$ search from $b\bar{b}$-associated production is not sensitive enough to provide a constraint).
In both cases, the coupling $y_{\chi}$ is fixed at each point to match the DM relic density.
In addition, Figure~\ref{PLOT1} shows the present and projected to $300$ fb$^{-1}$ (when possible) constraints on the dark portal discussed in the previous 
section: the exclusion from CMS/LHCb 8 TeV $B_s\to\mu^+\mu^-$ measurements (red), 
the multi-jet $+ \slashed{E}_{T}$ (black), and the $\bar{b} b$-associated production of $H_0 \to \tau \tau$ (blue) and $a \to \tau \tau$ (brown).
For the latter two, we use {\sc SusHi} to obtain the NNLO $H_0,\,a$ production 
cross section in association with $\bar{b} b$ at $13$ TeV LHC\footnote{We note that by performing the analysis of mono-$Z$ and our
$\bar{b} b Z( \rightarrow \ell\ell) + \slashed{E}_{T}$ signature at LO, as compared to $H_0,\,a \to \tau \tau$ at NNLO, 
we are being conservative by underestimating the constraining power 
of the former two signatures.}.  
We note that $\bar{t}t + \slashed{E}_{T}$ and $\bar{b}b + \slashed{E}_{T}$ are not sensitive enough to provide a constraint in 
Figure~\ref{PLOT1}. 

\begin{figure}[h!]
\begin{center}
\includegraphics[width=0.485\textwidth]{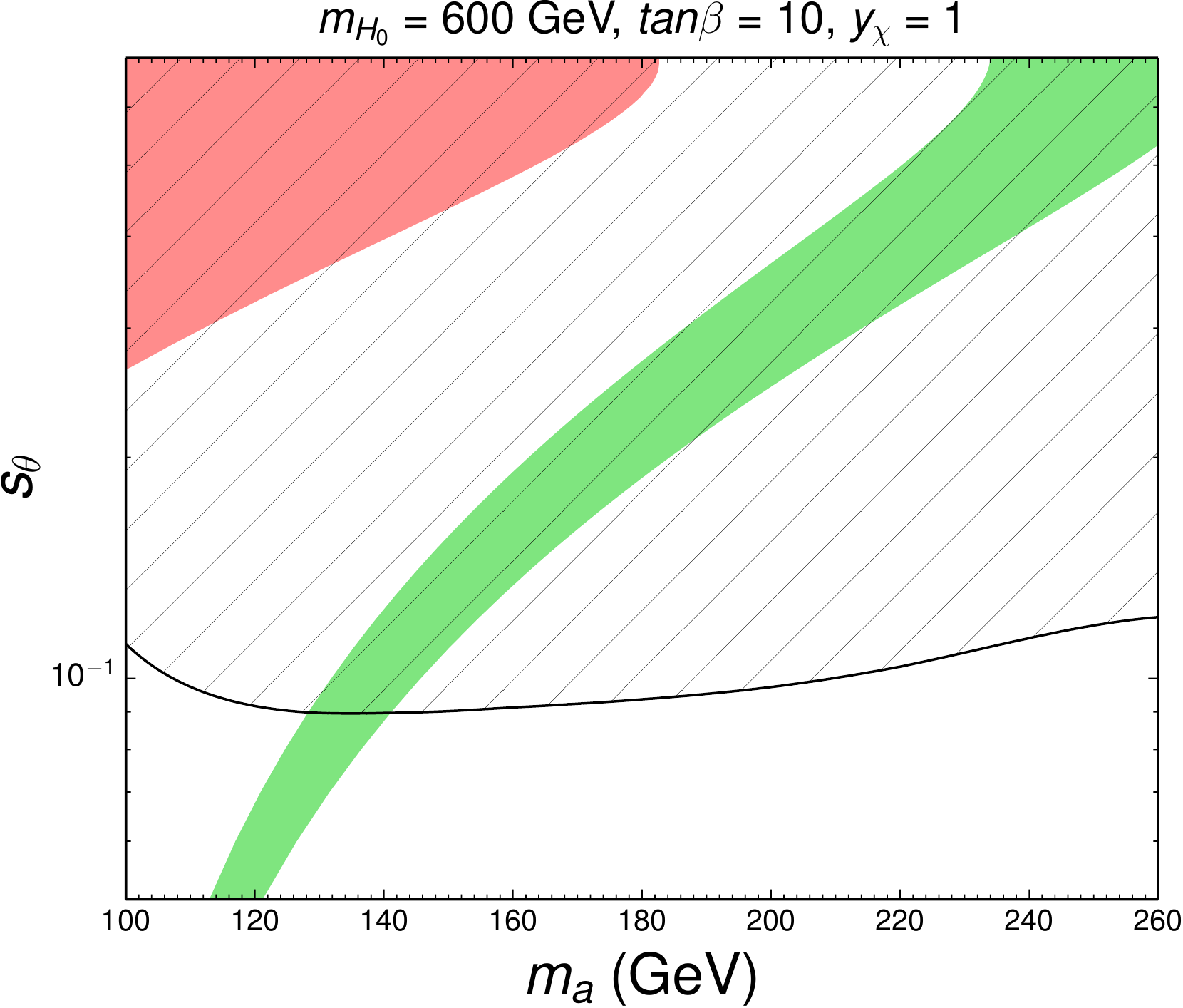}
\caption{\small 95\% C.L. sensitivity of the proposed search $pp\rightarrow b\bar{b}\,\ell\ell+\slashed{E_T}$ with $\mathcal{L}=300$ fb$^{-1}$ (dashed region)
in the ($m_{a},\,s_{\theta}$) plane for $m_{H_0} = 600$ GeV, $\mathrm{tan}\beta = 10$, $y_{\chi} = 1$. The red region is excluded by  
$B_s\to\mu^+\mu^-$, while the green band yields the observed DM relic density. The DM mass is $m_{\chi} = 45$ GeV.}
\label{PLOT2}
\end{center}

\vspace{-3mm}

\end{figure} 

As Figure~\ref{PLOT1} highlights, the ATLAS mono-$Z$ search will be able to probe the $\mathrm{tan}\beta \lesssim 5$ region (for GF production), 
while $B_s\to\mu^+\mu^-$ and the projected $H_0 \to \tau \tau$ combined could strongly constrain the very high $\mathrm{tan}\beta$ 
region ($\mathrm{tan}\beta > 10$ for $m_{H_0} = 600$ GeV, $\mathrm{tan}\beta > 20$ for $m_{H_0} = 800$ GeV); meanwhile, 
the proposed search $pp\rightarrow \bar{b} b \,\ell^+ \ell^- +\slashed{E_T}$ would yield access to the intermediate $\mathrm{tan}\beta$ region, probing 
also values of the mixing down to $s_{\theta} \sim 0.1$.

We note that in the above analysis, we have fixed $\Gamma_{H_0 \to a a} = 0$ (as can be done by an appropriate choice of the soft $\mathbb{Z}_2$ symmetry 
breaking term in the 2HDM scalar potential, see e.g.~\cite{Bauer:2017ota}). A non-vanishing $\Gamma_{H_0 \to a a}$ would weaken the constraints 
from mono-$Z$, our new signature $pp\rightarrow \bar{b} b \,\ell^+ \ell^- +\slashed{E_T}$ and from $H_0 \to \tau \tau$, but would at the same time yield 
new avenues to probe the pseudoscalar portal. We do not consider this scenario here for simplicity.

\vspace{1mm}

Finally, in Figure~\ref{PLOT2} we show the various constraints and projected sensitivities discussed above in the ($m_a,\,s_{\theta}$) plane for a benchmark 
$m_{H^{\pm}} = m_{H_0} = 600$ GeV, $\mathrm{tan}\beta = 10$ and $y_{\chi} = 1$, together with the 
$\langle \sigma \mathrm{v}\rangle = (2-4)\times 10^{-26}{\rm  cm}^3/{\rm s}$ region where the observed DM relic density 
is obtained (green). This highlights the sensitivity of the proposed search to the parameter space region with the correct DM relic density 
(and favoured by the GCE excess) as compared to other experimental probes of the pseudoscalar portal to DM.  

\vspace{-4mm}

\subsection*{V. Conclusions}

\vspace{-4mm}


DM that interacts with the visible sector via a pseudoscalar mediator is an appealing scenario, naturally avoiding the 
limits from DM direct detection searches while generating a rich LHC phenomenology and yielding a possible explanation for the 
FERMI gamma ray Galactic Centre Excess. Generating a pseudoscalar coupling to SM fields in a consistent way implies the existence of 
additional BSM particles, as in theories with two Higgs doublets where the necessary coupling is naturally generated when the pseudoscalar mediator 
and that of the two-Higgs-doublet scenario mix. We have shown that such scenarios 
give rise to a new LHC DM search channel $\bar{b} b H_0, H_0 \rightarrow Z a (Z \rightarrow \ell^+ \ell^-, a \rightarrow \bar{\chi} \chi)$.
The final state with a leptonically decaying $Z$ boson, $b$-tagged jet(s) and large $\slashed{E_t}$ has not been explored 
yet at the LHC in the DM context.


We find that a large region of parameter space which gives the observed DM relic abundance (yielding at the same time an explanation for the Galactic Centre Excess) 
can be explored using the proposed search, showing in particular that it can reach a wide region of parameter space that cannot be probed by other means, 
notably $B_s \rightarrow \mu^+ \mu^-$ decays, heavy Higgs ($H_0$) decays into tau-lepton pairs, and mono-$Z$ searches. 
This novel search can thus be very valuable in probing pseudoscalar portal DM scenarios at the LHC.

\vspace{8mm}

\begin{center}
\textbf{Acknowledgements} 
\end{center}

\vspace{-1mm}

\begin{acknowledgments}
J.M.N. thanks Seyda Ipek for a very insightful talk that inspired this project, as well as David Cerdeno and Richard Ruiz for useful discussions. 
P.T. and M.F. thank Bobby Acharya for helpful discussions.
M.F., J.M.N. and P.T. are supported by the European Research Council under the
European Union’s Horizon 2020 program (ERC Grant Agreement no.648680 DARKHORIZONS).  The work of MF was supported partly by the STFC Grant ST/L000326/1.
\end{acknowledgments}


\end{document}